\documentclass[10pt,final,journal,letterpaper]{IEEEtran}
\usepackage[utf8]{inputenc}
\usepackage[font=small,labelfont=bf]{caption}
\usepackage[margin=1in]{geometry}
\usepackage[export]{adjustbox}
\usepackage{bm, cite,floatrow,wrapfig,longtable,hyperref,bm,float,courier,enumitem,graphicx,wrapfig,multicol,multirow,amsmath,lscape,amsfonts,amssymb,setspace,gensymb,graphicx,fancyhdr,scrextend,courier,siunitx,subcaption,amsmath,lscape,amsfonts,amssymb,setspace,gensymb,graphicx,caption,listings,xcolor}

\definecolor{navyblue}{rgb}{0,0,0.4}
\definecolor{blueviolet}{rgb}{0.4,0,0.4}

\newenvironment{Table}
  {\par\medskip\noindent\minipage{\linewidth}}
  {\endminipage\par\medskip}
  
\title{Efficiency-optimized design of PCB-integrated magnetorquers for CubeSats}
\author{Nicholas J. Sorensen \thanks{Final Manuscript received by ITAES May 5, 2021.}%
\thanks{N. J. Sorensen is with the University of Alberta.//}}

\begin{document}

\setlength{\abovedisplayskip}{9pt}
\setlength{\belowdisplayskip}{9pt}

\setlength{\parindent}{0pt}

\maketitle

\begin{abstract}
CubeSats are miniature satellites used to carry experimental payloads into orbit, where it is often critical to precisely control their attitude. One way to do this is through the use of magnetorquers, which can be integrated into PCBs. This technique saves considerable space and capital when compared with more common torque-rod magnetorquer systems. Here we derive a method of analyzing different PCB-integrated magnetorquer geometries, parametrizing them such that the magnetic moment and efficiency are optimized. Furthermore, by modulating the trace width, the trace number, and other electrical characteristics of the magnetorquer coil, this paper optimizes the generated magnetic moment. Both constant voltage and constant current sources are analyzed as inputs. These optimizations are then simulated in COMSOL for multiple geometries, and it is found that there exists an optimal geometry, given a specified power dissipation. Simulations verify the general trend and maxima of these derivations, barring small, consistent re-scaling in the magnitude of the coil resistance. It is also found that these PCB-integrated magnetorquers provide a sufficient alternative to commercial coil magnetorquers - particularly in volume-restricted configurations. This study extends such analysis to larger CubeSat configurations, and finds that these larger implementations increase magnetorquer efficiency. Optimizations for common PCB-implementable geometries on small satellites are tabulated in the Appendix.  \\
\end{abstract}


\section{Introduction}
\label{Introduction}
\IEEEPARstart{C}{ube satellites}, also known as CubeSats, are small satellites frequently built due to their low cost and high modularity, and are commonly used to carry small, experimental payloads into orbit \cite{Ali2012, Musiab2016}. They are typically categorized by the number of units that they are comprised of, where one unit (1U) is a $(10 \times 10 \times 10$) cm$^3$ module. Commonly, CubeSats are 1U, 2U, 3U, or even 6U, and one of the common challenges in operating these satellites is controlling the CubeSat's attitude, or its orientation in space. This is done using some sort of attitude determination and control system (ADCS) \cite{Alminde2003,Ali2012,steyn1995,Musiab2016,Chen1999,Aman2015,CalPoly2009,Bellini2014,Mukhtar2016,Lappas2002,Francois-Lavet2010, Mughal2020, Ali2018-2}. \\

A typical commercial-off-the-shelf (COTS) ADCS system is comprised of a combination of reaction wheels and magnetorquer rods distributed on three orthogonal axes. These COTS magnetorquers nominally consume about 200 mW and produce a moment about 0.2 Am$^2$ \cite{CubeSpaceSatelliteSystemsRFPtyLtd2019}; however, these systems are commonly very bulky and consume a large percentage of the CubeSat's volume budget \cite{steyn1995}. As space is often at a premium on CubeSats - their smallness can often be restrictive - it is expedient to reduce the size of such components. \\

One way to make these systems smaller is to collapse the 3D magnetorquer rods into pseudo-2D PCB-integrated magnetorquers, considerably reducing their bulk. There is a history of the implementation of these embedded magnetorquers, and some recent investigations have also studied the optimization of PCB-integrated magnetorquers \cite{Ali2012, Mukhtar2016, Musiab2016, Aman2015,Ali2020, Mughal2020, Grau2017, Grau2019}. This paper differentiates itself by not only examining discretized torquing efficiency in a novel way, but by also exploring both constant voltage and constant current bases, while previous publications have typically only focused on a single electrical basis. Here, we present analysis that examines the interdependencies of geometrical parameters, and frame it in order to ease design and implementation. It is the goal of this paper to formulate a methodology to optimize such magnetorquers for a variety of CubeSat sizes (most publications focus on PC104 or 1U panels), and to find if they provide a good alternative to COTS magnetorquers.  Ultimately we compare their effective magnetic dipole moments and the amount of power they consume. \\

Magnetorquers are electronic devices that consist of coils of wire through which current runs, inducing a magnetic dipole moment \cite{steyn1995,CalPoly2009,Bellini2014}. This dipole interacts with the Earth's magnetic field, torquing the satellite. Given an external magnetic field $\bm{B}$, and a magnetic dipole moment $\bm{m}$, the torque on the satellite is given by:
\begin{equation}
\label{eq:torque}
\bm{\tau} = \bm{m} \times \bm{B}. 
\end{equation}
To increase the torque imparted onto the CubeSat, it is necessary to increase the magnetic moment, and this is typically facilitated by increasing the current passing through the coils, or by increasing the number of coils. This is easily established in a 3D system, but in a pseudo-2D system like that on a printed circuit board (PCB), coils cannot transversely overlap, and the thickness of the coil is relatively set. Hence, it is the purpose of this study to find an optimal design for a magnetorquer, producing a maximal magnetic dipole moment.\\ 

After defining the spatial restraints, this system is optimized by directly investigating the magnetic dipole moment, the trace-width-dependent and trace-number-dependent efficiency, and the power-isometric relations to each of these parameters. The analyzes of the efficiency and moment should return identical optimized solutions, but, by exploring both, any similar conclusions are bolstered in their validity.\\

\section{Design}
\label{Design}
\subsection{Geometry}
\label{Geometry}
\begin{figure}[b]
\centering
\includegraphics[width=0.9\linewidth]{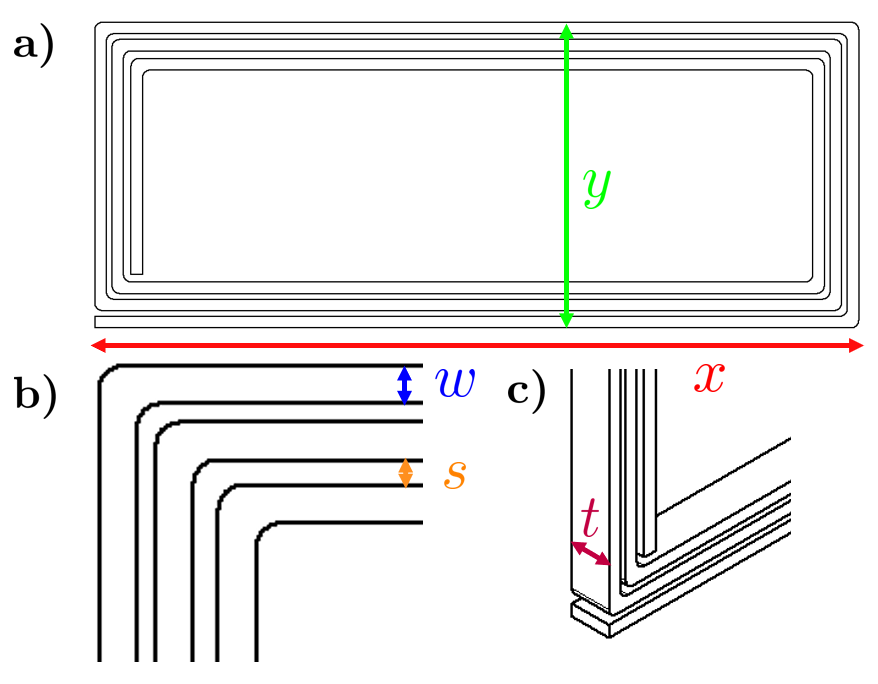} 
\captionof{figure}{Geometrical and dimensional definitions of a trace magnetorquer. Note that $t$ is expanded in scale from its actual size $(t \approx \SI{0.07}{mm} \leftrightarrow 2$ oz/ft). $x$ and $y$ are the width and height restraints, respectively, and $w$ and $s$ are the individual trace width and trace separation. $N$, unlabeled, is the number of traces; here $N=3$. a) Full, simplified geometry. b) Zoomed corner of geometry. c) Projected view of trace.}
\label{Mag_dim}
\end{figure}

The magnetorquer discussed in this report is designed to be implemented on a PCB, which means that said designs must be implementable in design software; hence, for simplicity, the coil takes a rectangular shape with dimensions as seen in Figure \ref{Mag_dim}. As well, the thickness, $t$, of the trace is commonly listed in units of oz/ft and is discretized; this analysis assumes a commonly available thickness of 2 oz/ft (0.07 mm). The trace corners are rounded to lessen impedance, and in fact, it may be best to use small $45\degree$ corners, in addition to the rounding, but that is reserved for an additional investigation. The other dimensions are essentially continuously variable and depend on the specifications of the satellite. \\

\subsection{Magnetic Moment}
\label{Design-Magnetic-Moment}
Physical systems tend towards orientations that are less energetic, and electromagnetic systems are no different: magnetic dipoles experience a torque when exposed to an external magnetic field in order place the system in a less energetic state. To induce a magnetic dipole moment, one can run a current $I$ around a loop, and its vector potential, $\bm{A}$, can be represented as a multipole expansion. As there exists no magnetic monopole the dipole term is dominant in the multipole expansion:
\begin{equation}
\label{eq:vec_pot}
\bm{A}_{\text{dip}}(\bm{r}) = \frac{\mu_0}{4 \pi}\frac{\bm{m}\times\bm{\hat{r}}}{r^2} \nonumber
\end{equation}
where $\bm{m}$ is the magnetic dipole moment \cite{griffiths2019}.\\
\begin{equation}
\label{eq:mag1}
\bm{m} = I\int \text{d}\bm{S} = I\bm{a} \nonumber
\end{equation}
given a current loop of area $\bm{a}$. For a volume current density, $\bm{J}$, the expression for $\bm{m}$ becomes:
\begin{equation}
\label{eq:mag2}
\bm{m}=\frac{1}{2}\int(\bm{r}\times\bm{J})\cdot\text{d}\tau
\end{equation}
where $\bm{r}$ is a position vector originating at the center of the geometry, and $\tau$ is a volume element. To maximize the dipole moment, one must therefore maximize the current present in the loop and the area encompassed by the current loop. It now becomes necessary to analyze this system under two different bases: constant voltage (CV), and constant current (CC) inputs.\\

\subsubsection{Constant Voltage}
\label{Design-Magnetic-Moment-Constant-Voltage}
This basis assumes that a constant potential is applied, such as those used on power buses on CubeSats (typically 3.3 V or 5.0 V). This potential is directly applied to one terminal of the coil, with the other terminal grounded. A simple resistor could be appended to specify the power consumed by the coil, acting as a voltage divider, but that would be mean power dissipation over a useless component. This is avoided. \\

Suppose a CV source of potential $V$ powers the coil. $\bm{m}_{\text{CV}}$ is then:
\begin{equation}
\label{eq:mag3}
\bm{m}_{\text{CV}} = \frac{V}{R}\left(\sum_{n=1}^{N}x_n y_n\right)\bm{\hat{n}}
\end{equation}
for the surface normal unit vector $\bm{\hat{n}}$, the individual lengths and heights of the loops $x_n$ and $y_n$ (m), respectively, referenced to the center of the trace, and the coil resistance, $R$ ($\SI{}{\Omega}$).
\begin{equation}
\label{eq:res}
R = \frac{\rho L}{t w}\left(1+\alpha(T-25\degree\text{C})\right)
\end{equation}
where $T$ is the temperature in $\degree$C, $L=\sum^N_{n=1}2(x_n+y_n)$ is the total length of the coil in m, and $\rho$ ($\SI{}{\Omega m})$ and $\alpha$ ((C$^{\degree})^{-1}$) are the resistivity and temperature coefficients, respectively. Note that $(1+\alpha(T-25\degree$C$))\approx 1$, for $T\approx 25 \degree$ C; this approximation will be used for the remainder of the derivation. The simulations also assume a temperature of 25$\degree$C. Combining Equations \ref{eq:mag3} and \ref{eq:res} we obtain:
\begin{equation}
\label{eq:mag4}
\bm{m}_{\text{CV}} =\frac{1}{2}\frac{V t w}{\rho}\left(\frac{\sum_{n=1}^{N} x_{n} y_{n}}{\sum_{n=1}^{N} x_{n}+y_{n}}\right) \hat{\bm{n}}.
\end{equation}
Here, and throughout the rest of the paper, $x_n$ and $y_n$ can be expressed in terms of $x$, $y$, $s$, $w$, and $n$: $x_n = x-w-2(n-1)(s+w)$; $y_n = y-w-2(n-1)(s+w)$ (note $x_n$ and $y_n$ reference the middle of the trace).\\

\subsubsection{Constant Current}
\label{Design-Magnetic-Moment-Constant-Current}
This basis assumes that a constant current is applied to one terminal of the coil. As in the CV model, a resistor could be appended as a current divider, but this would result in unwanted power dissipation, as before.\\

Suppose a CC source of current $I$ (A) circulates the coil. The resistance does not appear explicitly here, so the expression is much simpler than that in the CV case: 
\begin{equation}
\label{eq:mag5}
\bm{m}_{\text{CC}} = I\left( \sum_{n=1}^{N}x_n y_n \right) \hat{\bm{n}}.
\end{equation}

\subsection{Efficiency}
\label{Design-Efficiency}
One can also consider the power efficiency of the device. The designed magnetorquer is intended to have maximal efficiency, and efficiency is henceforth defined as $\eta:=\frac{P_{\tau}}{P_{\text{total}}}$ where $P_{\text{total}} = P_{\tau} + P_{\text{loss}}$, and $P_{\tau}$ is the power delivered to torquing the object, which is given by:
\begin{equation}
\label{eq:power1}
P_{\tau} = \bm{\tau} \cdot \bm{\omega} = (\bm{m} \times \bm{B})\cdot \bm{\omega}.
\end{equation}

Here, $\bm{\tau}$ is the torque (Equation \ref{eq:torque}), $\bm{B}$ is the external magnetic field, and $\bm{\omega}$ is the angular velocity of the system.\\

The other major source of power consumption in this system is through resistive losses ($P_{\text{loss}}$, given in further sections). Again, it becomes necessary to treat this system under two different bases: CV and CC.\\

\subsubsection{Constant Voltage}
\label{Design-Efficiency-Constant-Voltage}
As before, this model assumes that a constant potential is applied across the coil. The resistive losses in the system then total to $P_{\text{loss}} = \frac{V^2}{R}$ where $R$ is given by Equation \ref{eq:res}. Under CV Equation \ref{eq:power1} becomes 
\begin{equation}
\label{eq:power2}
P_{\tau}=\frac{V}{R}\bm{S}\left(\bm{\hat{{n}}}\times \bm{B} \right )\cdot \bm{\omega} \nonumber
\end{equation}
where $\bm{S}=\sum^N_{n=1}x_ny_n \bm{\hat{n}}$ is the sum of the areas of the coil. Hence, the efficiency of the system becomes 
\begin{equation}
\label{eq:eff1}
\eta_{\text{CV}} = \frac{\bm{S}\left(\bm{\hat{{n}}}\times \bm{B} \right )\cdot \bm{\omega}}{\bm{S}\left(\bm{\hat{{n}}}\times \bm{B} \right )\cdot \bm{\omega}+ V}.
\end{equation}   
To simplify further, let us consider a nominal application: in a CubeSat low earth orbit (LEO) mission, $B\approx 10^{-5}$ T \cite{MagField}, $S\approx1$ $\text{m}^2$, $\omega \approx 1 $ rad/s, and ideally assuming that $\bm{\hat{n}}$ is orthogonal to $\bm{B}$, and that the satellite is spinning axially parallel to $\bm{\hat{{n}}}\times \bm{B}$, we find that:
\begin{equation}
\label{eq:approx}
|\bm{S}\left(\bm{\hat{{n}}}\times \bm{B} \right )\cdot \bm{\omega}| \approx 10^{-5} << 5 \approx V. \nonumber
\end{equation}
We can therefore make the good approximation that
\begin{equation}
\label{eq:eff2}
\eta_{\text{CV}} \approx \frac{\bm{S}}{V}\left(\bm{\hat{{n}}}\times \bm{B} \right )\cdot \bm{\omega}. \nonumber
\end{equation}
As $\bm{\omega}$ and $\bm{B}$ are independent of this investigation, a proportionality is studied (scaled CV efficiency, $E_{\text{CV}}$):
\begin{equation}
\label{eq:eff3} 
\eta_{\text{CV}} \propto E_{\text{CV}} := \frac{S}{V} = \frac{1}{V}\sum_{n=1}^{N}x_n y_n.
\end{equation}

\begin{figure}[!t]
\centering
\includegraphics[width=\linewidth]{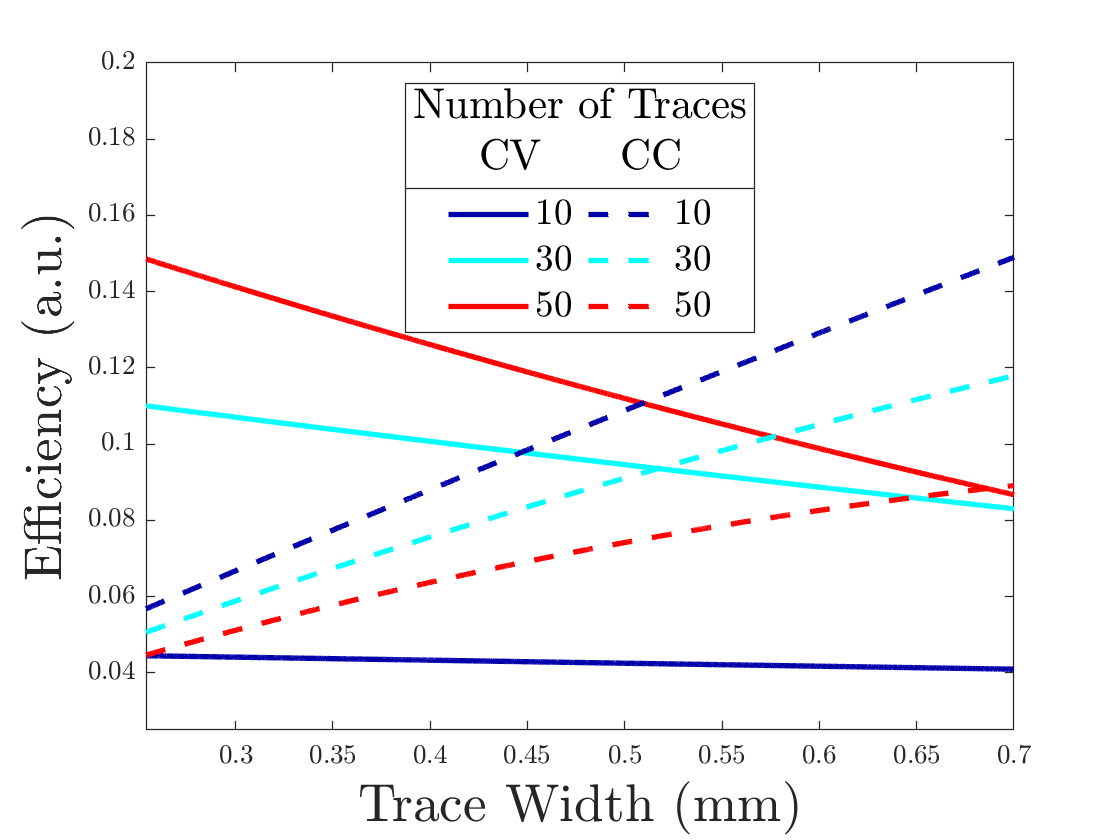}
\captionof{figure}{Given a fixed $x$, $y$, $\rho$, and (small) s, the efficiency, $\eta$, decreases as trace width increases for CV. For CC, however, the efficiency increases as trace width increases. Note that these efficiencies assume a spin rate of $\omega = 1$ Hz, $B = 1$ T, with $\bm{B}$ oriented in a direction orthogonal to $\bm{m}$. These are not realistic in a space environment, and efficiency would be many orders of magnitude lessened (though the same trends remain; see Equations \ref{eq:eff3} and \ref{eq:eff7}).}
\label{E_trace_CC_CC}
\end{figure}

By expanding $x_n$ and $y_n$ we get
\begin{equation}
\label{eq:eff4}
\begin{split}
\eta_{\text{CV}} \propto E_{\text{CV}}=\frac{1}{V}\sum_{n=1}^{N}\left[(x-w-2(n-1)(s+w))\cdot\ldots \right. \\
\left. (y-w-2(n-1)(s+w))\right]. \nonumber
\end{split}
\end{equation}
This evaluates to an analytic function which increases as $x$ and $y$ increase, and as $s$ and $w$ decrease (see Figure \ref{E_trace_CC_CC}). \\
\subsubsection{Constant Current}
\label{Design-Efficiency-Constant-Current}
As before, this model assumes that a constant current is applied to one terminal of the coil. The resistive losses then total to $P_{\text{loss}}=I^2R$ where $R$ is given by Equation \ref{eq:res}. For this system, 
\begin{equation}
\label{eq:power3}
P_{\tau} = I\bm{S}\left(\bm{\hat{{n}}}\times \bm{B} \right )\cdot \bm{\omega}. \nonumber
\end{equation}
The efficiency of the system becomes:
\begin{equation}
\label{eq:eff5}
\eta_{\text{CC}} = \frac{\bm{S}\left(\bm{\hat{{n}}}\times \bm{B} \right )\cdot \bm{\omega}}{\bm{S}\left(\bm{\hat{{n}}}\times \bm{B} \right )\cdot \bm{\omega}+ IR}. \nonumber
\end{equation}
Note that unlike Equation \ref{eq:eff1}, this equation is directly dependent upon $R$ though we can make the same approximation (Equation \ref{eq:res}). CC efficiency now becomes:
\begin{equation}
\label{eq:eff6}
\eta_{\text{CC}} \approx \frac{Stw}{I\rho L}\left(\bm{\hat{{n}}}\times \bm{B} \right )\cdot \bm{\omega}. \nonumber
\end{equation}
Again, as $\bm{\omega}$ and $\bm{B}$ are independent of this investigation, a proportionality is studied (scaled CC efficiency, $E_{\text{CC}}$):
\begin{equation}
\label{eq:eff7}
\eta_{\text{CC}} \propto E_{\text{CC}}=\frac{Stw}{I\rho L} = \frac{tw}{2\rho I}\frac{\sum_{n=1}^N x_n y_n}{\sum_{n=1}^N(x_n+y_n)}.
\end{equation}
As previously, it is also assumed that the current runs through the center of the trace, which is a good approximation except for at the corners of the loop. As in Equation \ref{eq:mag4}, $x_n$ and $y_n$ can be expressed in terms of $x$, $y$, $s$, and $w$. This evaluates to an analytic function which locally increases as $w$, $x$, and $y$ increase, and as $s$ decreases (see Figure \ref{E_trace_CC_CC}). Note that in $w$, this is the opposite effect as calculated for CV.\\

It should be said that this formulism is flexible in its geometry. By assuming the flow of current in the trace at its center, this derivation and the following one's can be adapted to any trace geometry, and it can be shown that
\begin{align}
\label{eq:effCV}
\eta_{\text{CV}} \propto E_{\text{CV}}&=\frac{1}{V}\sum_{n=1}^N \iint_{S} \text{d}S\nonumber\\
& = \frac{m_{\text{CV}}}{P_{\text{CV}}} 
\end{align}
\begin{align}
\eta_{\text{CC}} \propto E_{\text{CC}}&=\frac{tw}{\rho I} \left[ \sum_{n=1}^N \iint_{S} \text{d}S\right]\left[ \sum_{n=1}^N \int_{C} \text{d}s\right]^{-1}\nonumber\\
\label{eq:effCC}
& = \frac{m_{\text{CC}}}{P_{\text{CC}}}
\end{align}
where $P_{\text{CV}}$ and $P_{\text{CC}}$ are the corresponding CV and CC power consumptions, discussed in the following section, given by Equations \ref{eq:powerCV} and \ref{eq:powerCC}.
\subsection{Power Consumption}
\label{Design-Power-Consumption}
It may seem that this subject is being treated with an overbearing formulism, but the interdependence of many of the variables is nuanced. One could define $w$ in terms of $N$ but that would strongly limit the combinations, and it is difficult, if not impossible to separate the variables in the expression for power or resistance; the most efficient system is difficult to quantify. Fortunately, one's geometric requirements define a few of the variables ($x$, $y$, $s$, $V$, $I$) but many variables remain interdependent ($N$, $w$, $P_{\text{total}}$, $R$, $E$). \\

The previous derivation (Section \ref{Design-Efficiency}) is slightly misleading in that it produces the most efficient coil for systems with a constant number of coil turns, but not with a constant power consumption. The power absorbed by the magnetorquer system is, as previously shown, equal to the power lost to heat (the power absorbed by the dipole moment is negligible). Magnetorquer systems are often parametrized not by the number of coils they have, nor their induced magnetic moment, but by their power consumption. Hence, it is the purpose of this section to derive the most efficient system by comparing those with isometric power draws. Both CV and CC systems are dependent upon the coil's resistance.
\begin{equation}
\begin{split}
\label{eq:res2}
R &= \frac{2\rho}{tw}\sum_{n=1}^{N}(x_n + y_n) \\ 
&=\frac{\rho N}{tw}[2(x+y-2w)-4(s+w)(N-1)]. \nonumber
\end{split}
\end{equation}
The power consumption for a CV coil then becomes
\begin{equation}
\label{eq:power4}
P_{\text{CV}} = \frac{V^2 tw}{\rho N}[2(x+y-2w)-4(s+w)(N-1)]^{-1}.
\end{equation}
Conversely, the power consumption for a CC coil is
\begin{equation}
\label{eq:power5}
P_{\text{CC}} = \frac{I^2\rho N}{tw}[2(x+y-2w)-4(s+w)(N-1)].
\end{equation}

\begin{figure*}[!t]
\normalsize
\begin{subfigure}{.45\textwidth}
\centering
\includegraphics[width=\linewidth]{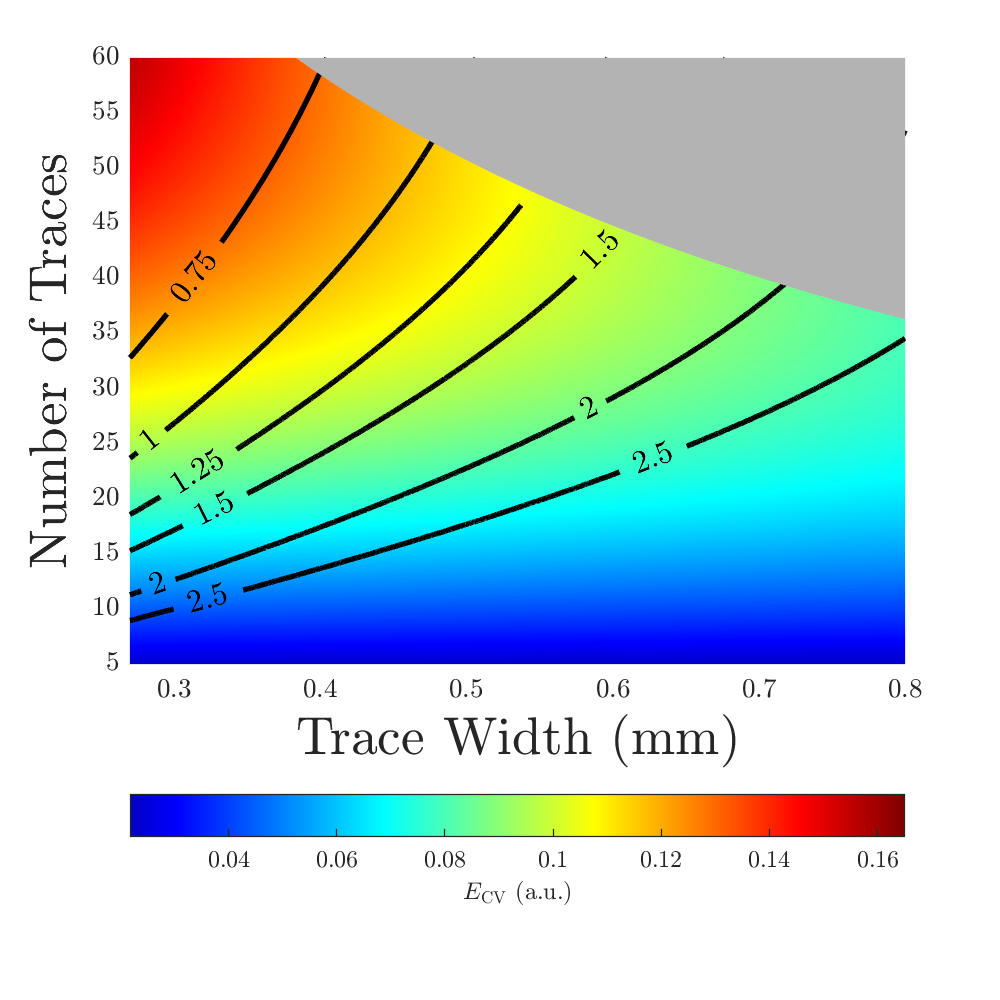} 
\captionof{figure}{Constant Voltage}
\label{E_power_a}
\end{subfigure}
\begin{subfigure}{.45\textwidth}

\centering
\includegraphics[width=\linewidth]{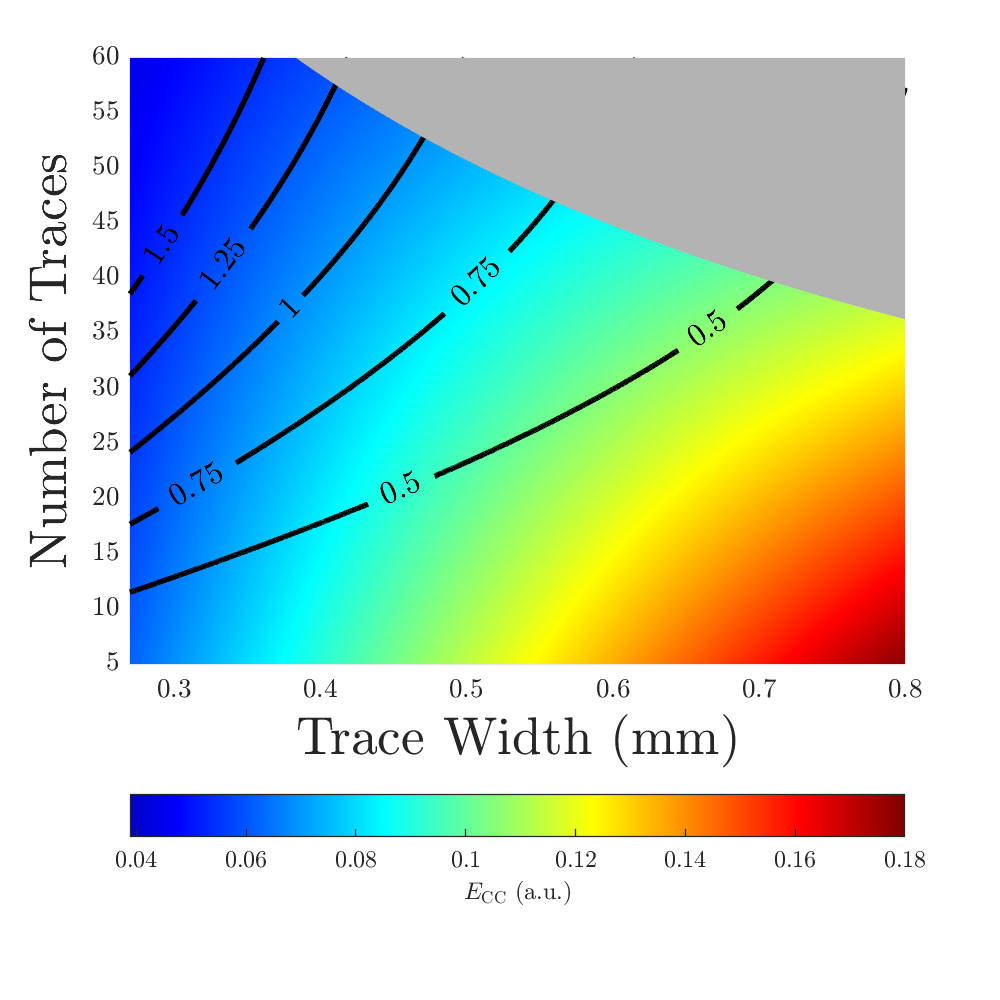} 
\captionof{figure}{Constant Current}
\label{E_power_b}
\end{subfigure}
\caption{Comparison plots of power consumed, and power efficiency (fixed parameters given in Table \ref{ta:parameters}) for (a) CV and (b) CC PCB magnetorquers. The surface plots denote the efficiencies of the systems given the number of coils and their trace width. There are isometric power contours with labeled powers in W. Above the greyed boundary, the data becomes non-physical. Also, note that the color scaling is inconsistent between plots. Reference Equations \ref{eq:eff3}, \ref{eq:eff7}, \ref{eq:power4} and \ref{eq:power5}.}
\label{E_power}
\end{figure*}

\begin{figure*}[!t]
\normalsize

\begin{subfigure}{.45\textwidth}
\centering
\centering
\includegraphics[width=\linewidth]{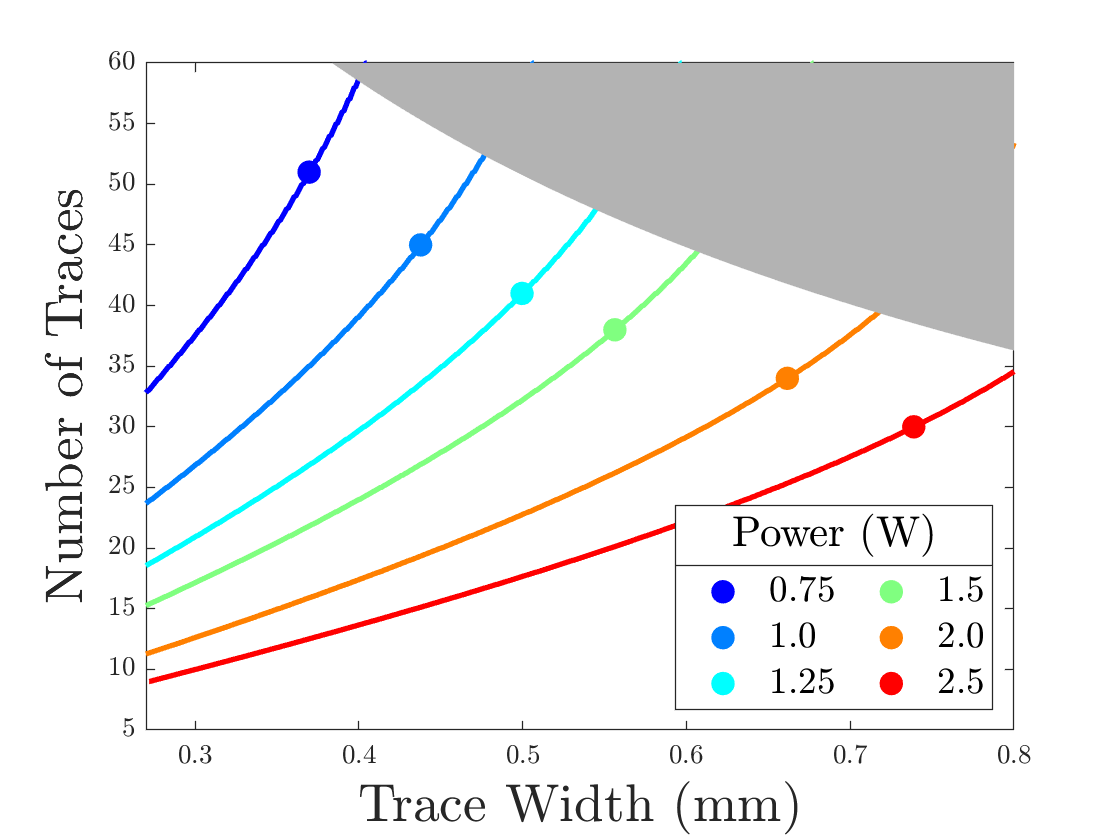} 
\captionof{figure}{Constant Voltage}
\label{E_max_a}
\end{subfigure}
\begin{subfigure}{.45\textwidth}
\centering
\includegraphics[width=\linewidth]{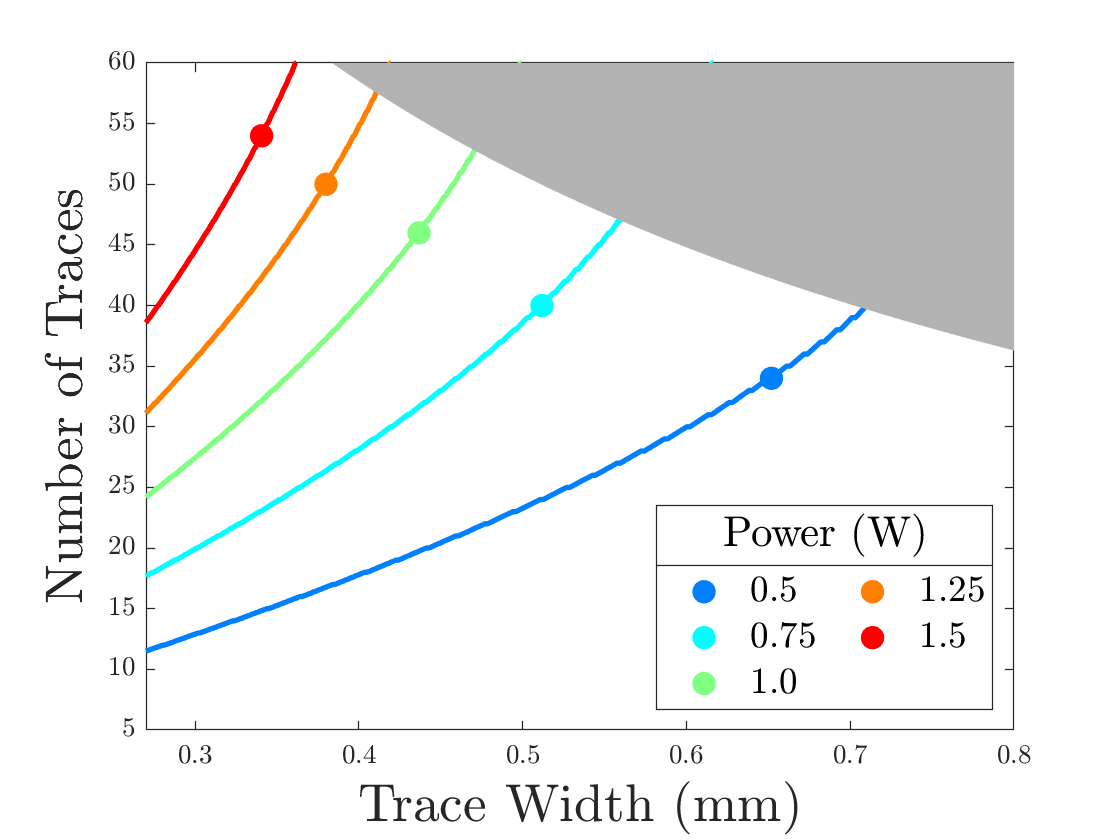} 
\captionof{figure}{Constant Current}
\label{E_max_b}
\end{subfigure}
\caption{Optimization of magnetorquer power efficiency given both (a) CV and (b) CC systems, introduced in Figure \ref{E_power}. Power is maximized at the scatter point on each isometric line. Note that only points below the grey boundary are physical. Further, note that the imaginary line that could connect the maxima is not smooth. As the number of traces is discrete, there is erratic stepping of the line of maxima. Reference Equations \ref{eq:eff3}, \ref{eq:eff7}, \ref{eq:power4} and \ref{eq:power5}.\\
\makebox[\linewidth]{\rule{\linewidth}{1.2pt}}}
\label{E_max}
\end{figure*}
Note that these equations have some limitations. Because $x$ and $y$ define the outer dimensions of the system, the coils only have room to expand inwardly. Defining a dimensionless coefficient $0<\beta<1$, we can define the max width of the stacked coil $W = \frac{\beta}{2}y$, given that $y\leq x$. $\beta$ is used here to provide a measure of fullness should some geometrical restrictions apply. The maximum number of turns in the coil given a trace width becomes:
\begin{equation}
\label{eq:N}
N = \bigg\lfloor\frac{\frac{\beta}{2}y + s}{w + s}\bigg\rfloor.
\end{equation}
This maximum limit is present in Figures \ref{E_power}-\ref{m_max} as the boundary of the plot and the grey upper block. Any data plotted within the grey area is non-physical.\\
Now the task becomes maximizing efficiency (or moment) while matching that efficiency (or moment) to a specific power consumption. One cannot express the efficiency, $E$, directly in terms of the power consumed, $P$, but one can solve the system numerically (see Sections \ref{sec:EffOpt}, and \ref{sec:MMOpt}).\\

As before for efficiency, the power consumption can be given for a general geometry:
\begin{figure*}[!t]
\normalsize
\begin{subfigure}{.45\textwidth}
\centering
\includegraphics[width=\linewidth]{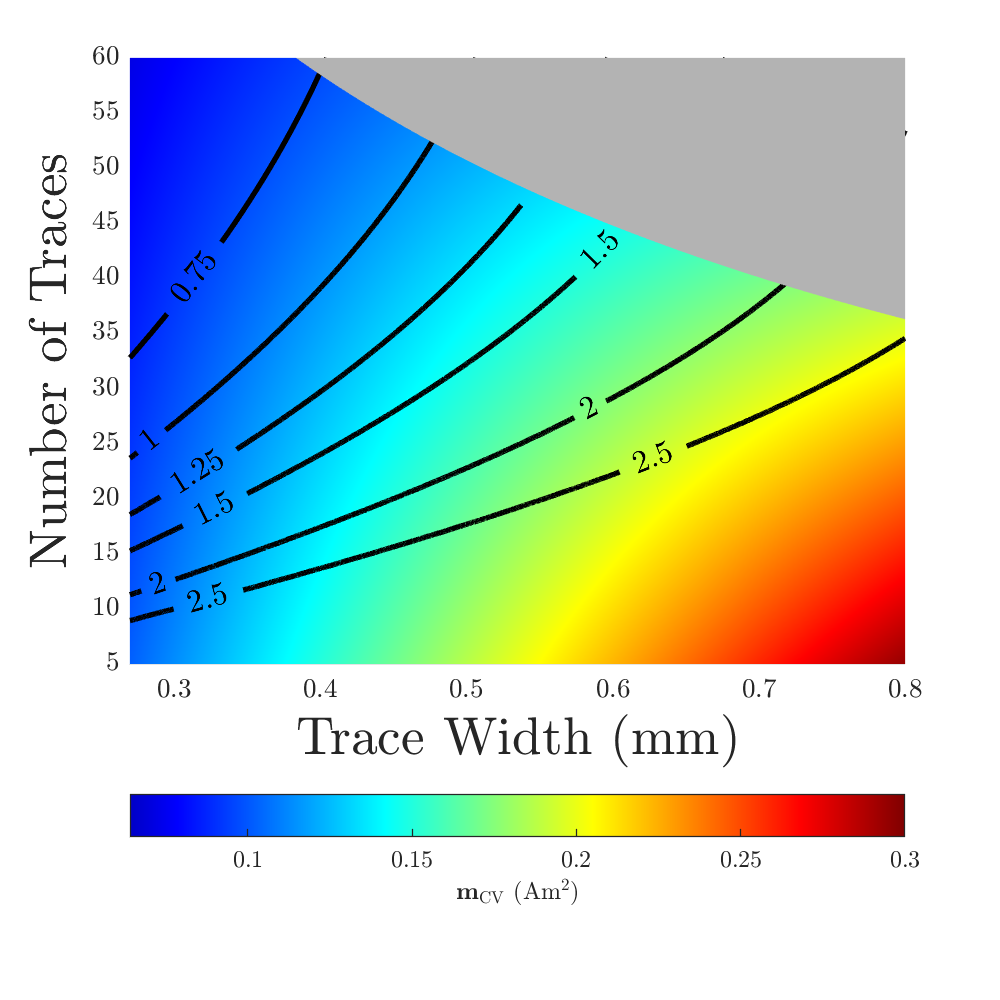} 
\captionof{figure}{Constant Voltage}
\label{m_power_a}
\end{subfigure}
\begin{subfigure}{.45\textwidth}

\centering
\includegraphics[width=\linewidth]{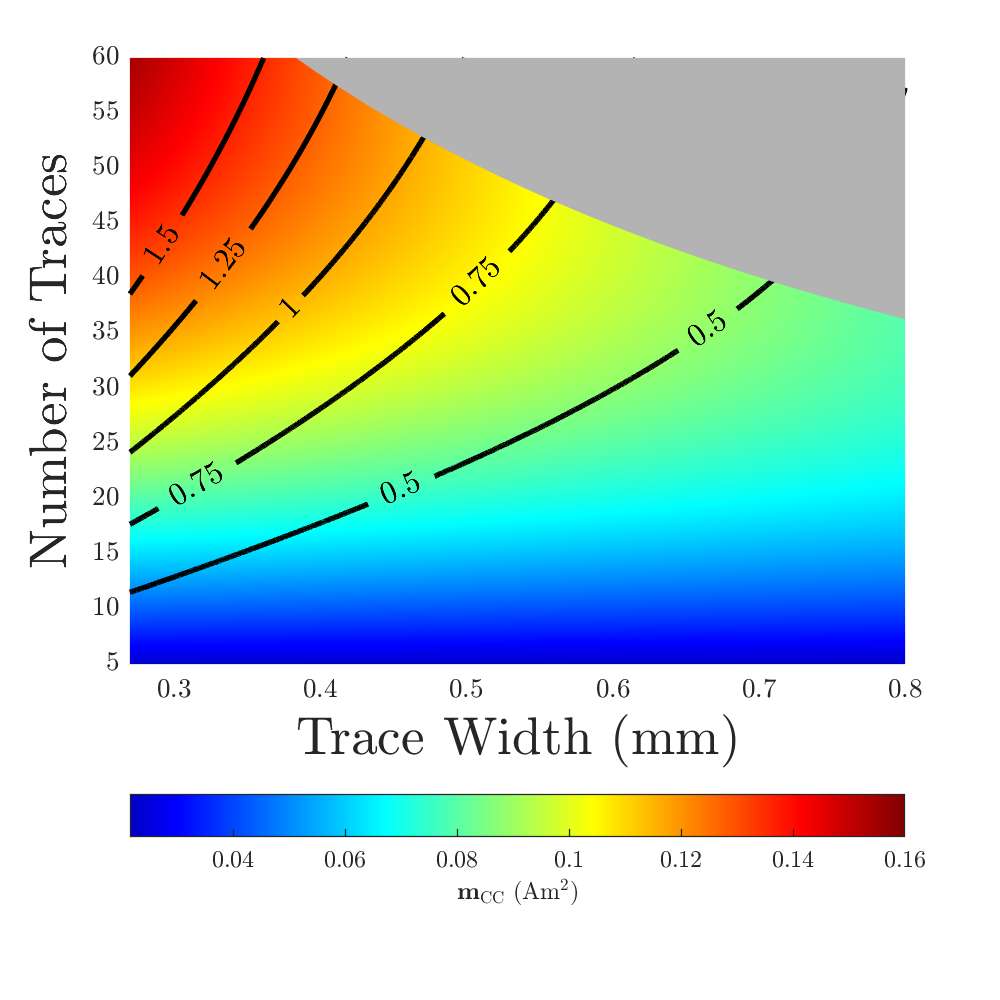} 
\captionof{figure}{Constant Current}
\label{m_power_b}
\end{subfigure}
\caption{Comparison plots of power consumed, and magnetic moment (fixed parameters given in Table \ref{ta:parameters}) for (a) CV and (b) CC PCB magnetorquers. The surface plots denote the magnetic moments of the systems given the number of coils and their trace width. There are isometric power contours with labeled powers in W. Above the greyed boundary, the data becomes non-physical. Also, note that the color scaling is inconsistent between plots. Reference Equations \ref{eq:mag3}, \ref{eq:mag5}, \ref{eq:power4} and \ref{eq:power5}.}
\label{m_power}
\end{figure*}

\begin{figure*}[!t]
\normalsize
\begin{subfigure}{.45\textwidth}
\centering
\centering
\includegraphics[width=\linewidth]{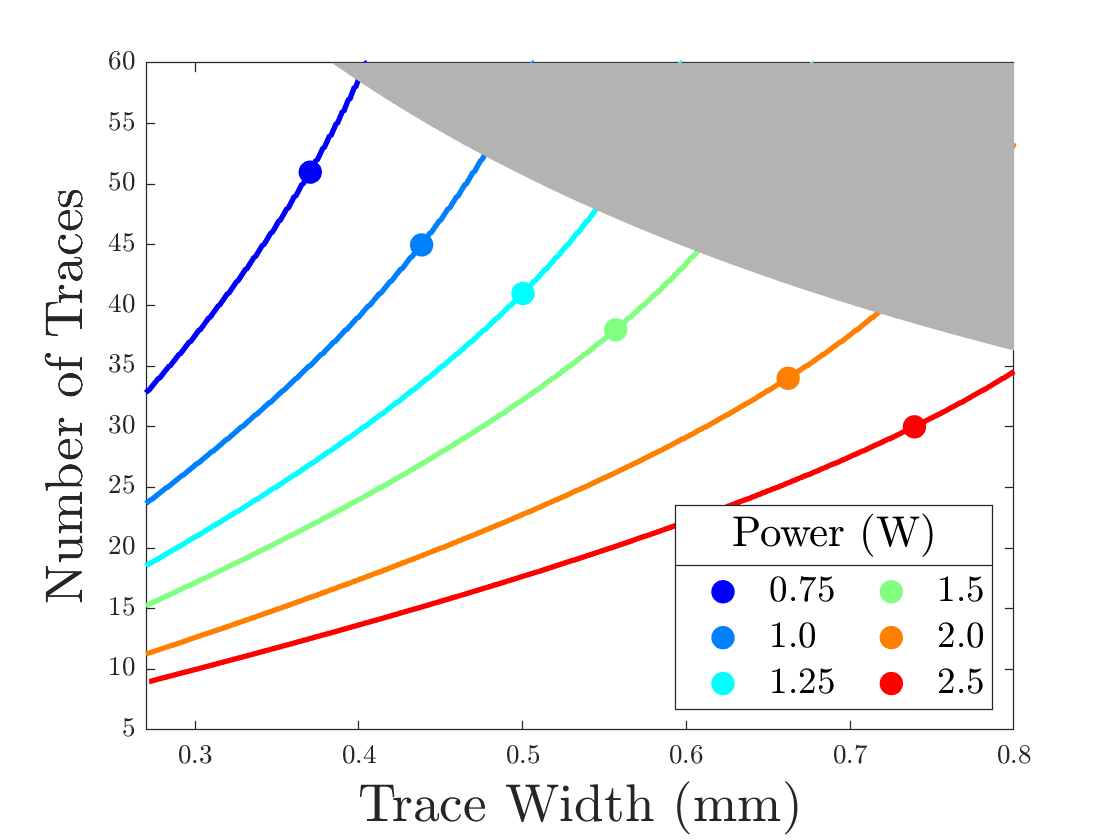} 
\captionof{figure}{Constant Voltage}
\label{m_max_a}
\end{subfigure}
\begin{subfigure}{.45\textwidth}
\centering
\includegraphics[width=\linewidth]{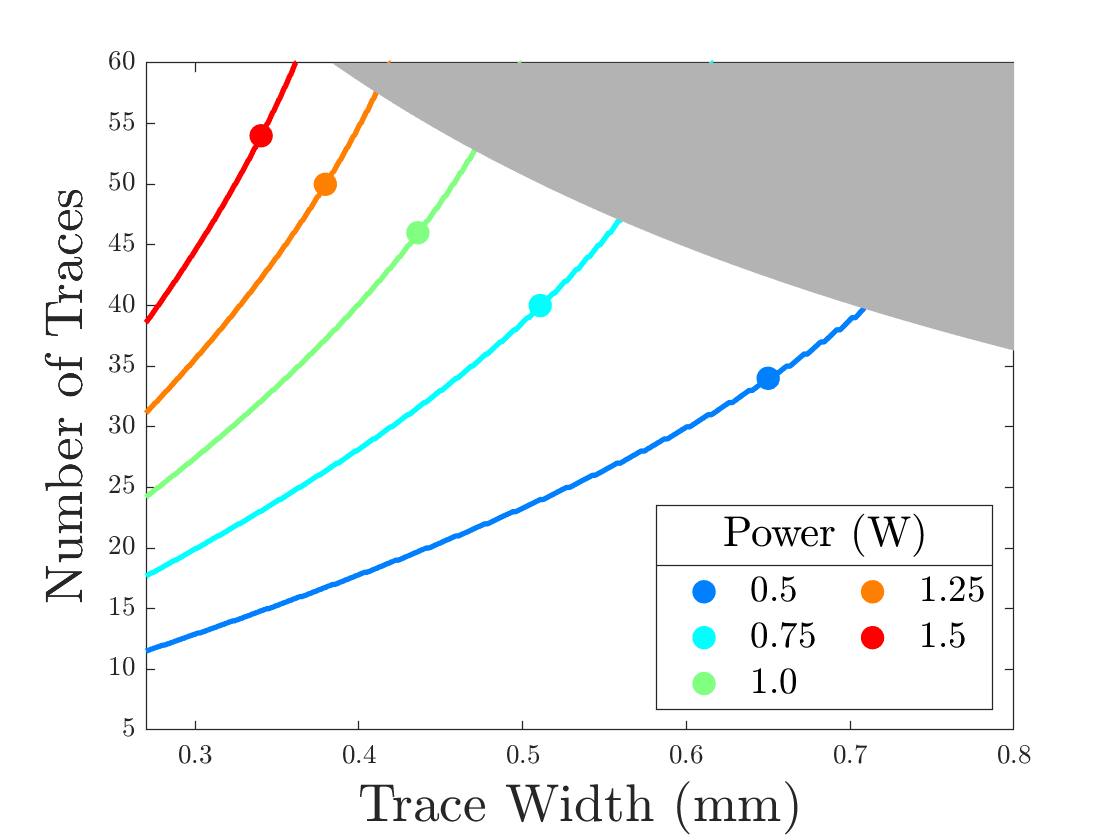} 
\captionof{figure}{Constant Current}
\label{m_max_b}
\end{subfigure}
\caption{Optimization of magnetorquer magnetic moment given both CV and CC systems, introduced in Figure \ref{m_power}. Moment is maximized at the scatter point on each isometric line. Note that only points below the greyed area are physical. Reference Equations \ref{eq:mag3}, \ref{eq:mag5}, \ref{eq:power4} and \ref{eq:power5}.\\
\makebox[\linewidth]{\rule{\linewidth}{1.2pt}}}
\label{m_max}
\end{figure*}

\begin{align}
\label{eq:powerCV}
P_{\text{CV}} = \frac{V^2 t w}{\rho}\left[ \sum_{n=1}^N \int_{C} \text{d}s\right]^{-1}.\\
P_{\text{CC}} = \frac{I^2\rho}{tw} \sum_{n=1}^N \int_{C} \text{d}s.
\label{eq:powerCC}
\end{align}

\subsection{Efficiency Optimization}
\label{Design-Efficiency-Optimization}
\label{sec:EffOpt}

Now the task becomes to optimize efficiency given a specific power consumption. One can do this by plotting isometric power over-top of 2D efficiency plots. Then, by finding the maximum efficiency along said isometric power lines, one optimizes the system.\\ 

\subsubsection{Constant Voltage}
\label{Design-Efficiency-Optimization-Constant-Voltage}

In reference to Figure \ref{E_power}, to compare different systems (differing in their trace width and number), one can compare the lines and the corresponding powers. The more red-shifted each line becomes (with respect to the surface color), the more efficient that system is. One can then solve numerically where along this power-isometric line efficiency is maximized [as long as it remains below the grey boundary, which denotes physicality (Equation \ref{eq:N})].\\

In Figure \ref{E_max}, efficiency is maximized for each power line at the denoted point. As long as the solution for each line is beneath the grey boundary, it is physical. If the solution is above the boundary one must either increase $\beta$ to increase the coil maximum width, or sacrifice efficiency and use the parameters given at or below the intersection of the non-physical boundary and the power-isometric line.\\

\subsubsection{Constant Current}
\label{Design-Efficiency-Optimization-Constant-Current}

Similar to the CV system, we wish to solve along each power contour for the highest corresponding efficiency (see Figures \ref{E_power} and \ref{E_max}). Any non-physical solutions are be remedied similarly to the CV case: either raise $\beta$ or use the parameters given at or below the intersection of the grey area and the power contour. For many of these power-isometric systems, there is a physical local maximum at which efficiency is optimized.

\subsection{Magnetic Moment Optimization}
\label{Design-Magnetic-Moment-Optimization}
\label{sec:MMOpt}
Now we aim to maximize the moment, given a certain power consumption. In fact, maximizing the moment should reproduce the results for maximizing efficiency, given the same geometrical basis. \\

\subsubsection{Constant Voltage}
\label{Design-Magnetic-Moment-Optimization-Constant-Voltage}
One can now overlay a color plot of the moment with power contours as we did before for efficiency (see Figure \ref{m_power_a}). Then, by finding the maximum moment along each power line, one can again find the most efficient system (see Figure \ref{m_max_a}). Comparing CV Figures \ref{E_power_a} and \ref{m_power_a} we see that efficiency and moment seem to be inversely related (as shown by the inversed color palettes in both figures) with respect to trace width and the number of traces; furthermore, we see by comparing Figures \ref{E_max_a} and \ref{m_max_a} that the maxima are exactly the same for the efficiency and moment cases, which is expected. These two figures are nearly identical, though they represent two different data sets and two different approaches to system optimization. This improves the validity of the approach to magnetorquer optimization demonstrated in this paper. \\

\subsubsection{Constant Current}
\label{Design-Magnetic-Moment-Optimization-Constant-Current}
One can now overlay a color plot of the moment with power contours as we did before for efficiency and CV moment (see Figure \ref{m_power_b}). Then, by finding the maximum moment along each power line, one can find the most efficient system (see Figure \ref{m_max_b}). Comparing CC Figures \ref{E_power_b} and \ref{m_power_b} we see that efficiency and moment seem to be inversely related as in the CV case, but this is because of the efficiency's lack of direct dependence on $R$; furthermore, we see by comparing Figures \ref{E_max_b} and \ref{m_max_b} that the maxima are exactly the same for the efficiency and moment cases, which is as expected.\\

\section{Application and Simulation}
\label{Application-and-Simulation}

\begin{figure}
\centering
\includegraphics[width=\linewidth]{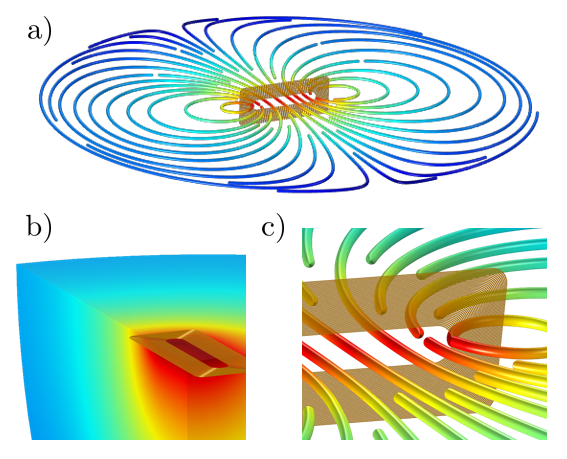} 
\captionof{figure}{Visualization of magnetic field lines and magnitude of the magnetic field produced by magnetorquer. Color plots display $\log_{10}(B)$ with higher values red-shifted.}
\label{mag_field_coil}
\end{figure}

To test the findings from previous sections, 4 geometries are studied, each tested at CV and at CC, and each designed to consume 1 W of power at CV (0.98 W at CC). The parameters for each geometry are given in Table 1 and Table 2, the latter of which shows a variety of different geometric parameters. The geometry is created in a CAD modeling program, then exported into COMSOL in which the boundary conditions are positioned and the physics defined. Each geometry is simulated using the \textit{Magnetic Fields }node in COMSOL, with an applied current or voltage using the \textit{Terminal} and \textit{Ground} subnodes/attributes. Once the simulation is run, a probe averages the magnetic flux density across a $4\times 4$ mm$^2$ area at the center of the coil geometry, and the magnetic dipole moment is computed using Equation \ref{eq:mag2}. See Figure \ref{mag_field_coil} for a visualization of the magnetic field facilitated through COMSOL.\\

\section{Results and Discussion}
\label{Results-and-Discussion}

\begin{figure}
\centering
\includegraphics[width=\linewidth]{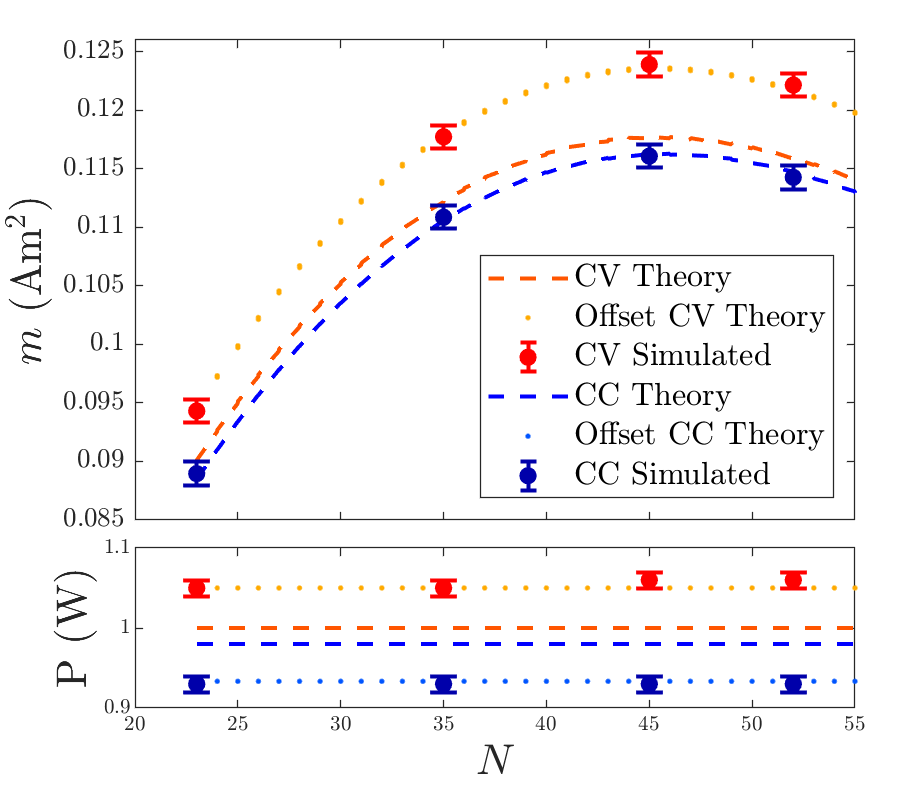} 
\captionof{figure}{Comparison of the calculated theoretical and simulated values for the magnetic moment of the magnetorquer.}
\label{m_comp}
\end{figure}

Comparing the theoretical and experimental data in Figure \ref{m_comp}, we have considerable agreement, with maxima occurring where expected. The simulated coils have somewhat different power consumptions when compared to the theory, but this is attributable to small differences in coil resistance (see Table \ref{ta:results}). By scaling the theoretical power consumptions of CV and CC by $(1.05\pm0.02)$ and $(1.050\pm0.005)^{-1}$, respectively, the data is fit within uncertainty of all data points. Considering that $P$ is directly proportional to $R$ in CC and inversely so in CV, this suggests that the theory over-predicts the value of $R$ by a factor of $(1.05\pm0.02)$. Strengthening this argument, let us consider the simulated magnetic moment: the CC data agrees within uncertainty of the theory without scaling, and the CV data agrees within uncertainty after scaling the theoretical magnetic moment by $(1.050\pm0.003)$. Considering Equations \ref{eq:mag3} and \ref{eq:mag5}, we must conclude that the theory over-estimates $R$ by a factor of 1.05, while accurately predicting the relationship between resistance and induced magnetic moment. In addition to the aforementioned non-idealities, the theory does not account for heterogeneous current flow or the curved corners, and this will affect the power consumption of the coil; in fact, the slightly curved corners and heterogeneous current flow may be responsible for the differing resistances, as well as the aforementioned influences of temperature. Discrepancy may also arise from moment calculation's assumption of 2D flow current flow, whereas the simulation uses a three-dimensional current flow calculation. These differences, however, should not, and do not influence the greater trends. The simulations agree with the maxima predicted by theory, and the ideal geometry can be derived from this formulism. \\

Comparing this kind of system to a commercially-available torquer rod, whose proportional efficiencies are given nominally by Equations \ref{eq:effCV}, and \ref{eq:effCC}, torquer rods have efficiencies of $\sim 1$ Am$^2$/W while our 2D systems are maximally efficient between $0.1\sim 0.5$ Am$^2$/W, depending on the geometry (see Table \ref{ta:optimizations}). Though these 2D systems cause a reduction in efficiency by about a factor of two, the lessened volume allocation in our pseudo-2D torquers provides great advantage. If a satellite configuration does not have much room for a bulky ADCS, the system we propose provides a good alternative, and our formulism provides the best way to optimize such a system. Should satellite teams desire to design PCB-integrated torquers into their system, this paper provides methods instrumental in creating an optimal system. Additionally, the methodology overviewed in this paper allows designers to implement PCB-integrated magnetorquers into larger panels than previously explicitly studied, increasing their efficacy.\\

\section{Conclusion}
\label{Conclusion}
This study sought to derive a method of optimizing the design of a magnetorquer capable of being integrated into a PCB. Optimizations were made for the both the magnetic dipole moment as well as for the power efficiency of the magnetorquer, whose relation to the magnetic dipole moment is defined and discussed. Parameters were varied, such as the trace width, the number of traces, as well as the consumed power. Four geometries were simulated in COMSOL to verify the theoretical basis, each calibrated to consume 1 W under a basis of constant voltage. It was found that while the derivations accurately predict the relationship between coil resistance and magnetic dipole moment, they also consistently overestimated the resistance of the loop by a factor of $(1.05\pm0.02)$.  Given geometric bounds and power consumption, the derived algorithms were shown to accurately calculate the appropriate parameters to define an optimized PCB-integrable magnetorquer system. Furthermore, the formulism discussed here agreed with and expanded upon previous work in the design and analysis of magnetorquers in greater detail, and examined the subject from two different electronic bases - something not previously done. Simulations showed the veracity of the detailed approach in actually designing optimized systems, and demonstrated that PCB magnetorquer systems are a good alternative to other torquing designs. Prototypical optimized designs for common CubeSat boards can be found in Table \ref{ta:optimizations} in the Appendix. 

\section{Future Work}
\label{Future-Work}
There is considerable impetus for designing pseudo-2D, PCB-integrable magnetorquer systems that could replace the bulky ADCS systems that dominate CubeSat builds today. Optimizing these systems involves not only optimizing the parameters that this study dealt with, but many other variables, such as copper thickness, PCB material, and, in particular, different boundary geometries, as well as considerations of temperature. Further studies should be done to rigorously explore these areas. Furthermore, the coils may adversely affect adjacent electronic components when they introduce a substantial magnetic field - this avenue should be explored. Finally, as the main goal of this study was to more rigorously define the formulism governing PCB magnetorquers, only simulations were explored; the implementation of these magnetorquers into a variety of differently sized panels (something that has not been previously demonstrated) motivates the testing these systems in a real environment, similar to that of in a LEO. Seldom are magnetorquer coils larger than 1U studied - one of the novelties of this paper - so the author is additionally interested in this avenue of exploration. The actual testing of these implementations does not easily fit within the scope of this paper, and is therefore reserved for future publication. Ideally, these real world tests are at the forefront of future work, and we are considering avenues to test the nature of these torquers in a real, experimental setting. 

\section{Acknowledgements}
\label{Acknowledgements}
This research was supported with funding provided by the Faculties of Science and Engineering at the University of Alberta. Additionally, we would like to thank Duncan G. Elliott, Michael Lipsett, Steven Knudsen, Ian R. Mann, Shirley Wang, and the Davis Group at the University of Alberta for useful conversations. We would also like to acknowledge the support of Natural Sciences and Engineering Research Council (NSERC).

\bibliographystyle{ieeetr}
\bibliography{library}

\begin{IEEEbiography}
    [{\includegraphics[width=1in,height=1.25in,clip,keepaspectratio]{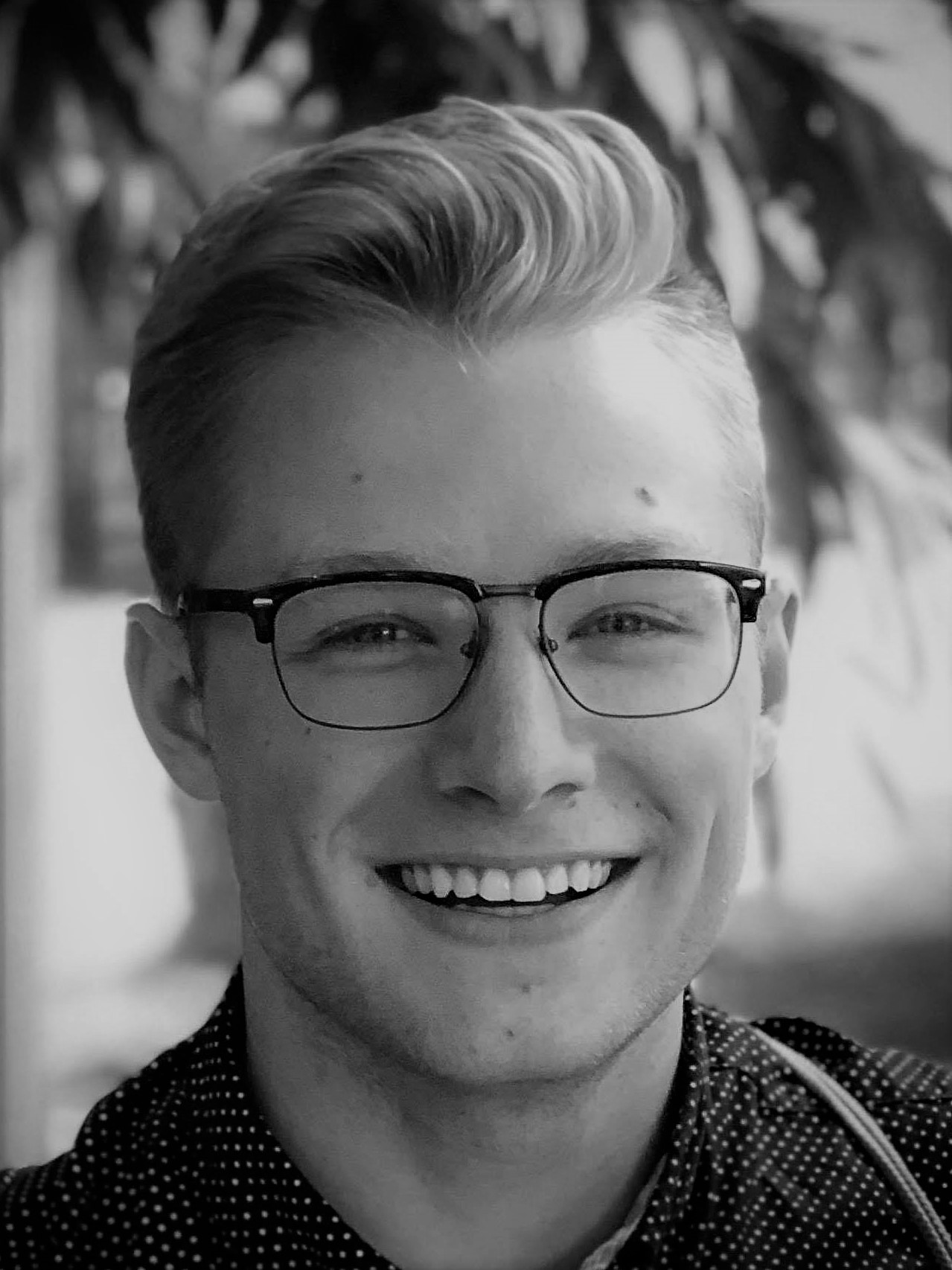}}]{Nicholas J. Sorensen}
received the B.Sc degree in engineering physics from the University of Alberta in 2021 and will be pursuing a M.Sc degree in the upcoming year. He is currently a leading systems and power designer within the AlbertaSat CubeSat Project and has also performed photonics and optomechanics research at the University of Alberta. His research interests include spacecraft and related electronic-control system development, photonics, and optoelectronics. 
\end{IEEEbiography}
\onecolumn

\appendix

\section{Appendix}
\label{Appendix}
\begin{multicols}{2}
\begin{Table}

\centering
\begin{tabular}{|c|c|c|}
\hline
\textbf{Variable} & \textbf{Value}   & \textbf{Unit} \\ \hline
$x$               & 200              & mm            \\ \hline
$y$               & 80               & mm            \\ \hline
$s$               & 0.254            & mm            \\ \hline
$t$               & 0.07             & mm            \\ \hline
$B_{\text{ext}}$  & $3.5\cdot 10^-5$ & T             \\ \hline
$V$               & 3.3              & V             \\ \hline
$I$               & 0.3              & A             \\ \hline
$\rho$            & $1.7\cdot10^-8$  & $\SI{}{\Omega m}$             \\ \hline
$\beta$           & $0.95$           & -             \\ \hline	
$r_{\text{fillet}}$           & 5           & mm             \\ \hline  
\end{tabular}
\captionof{table}{Fixed parameters for the simulated magnetorquer system. $r_{\text{fillet}}$ is the fillet radius for the corners of the coil.}
\label{ta:parameters}
\end{Table}

\begin{table}[H]

\centering
\begin{tabular}{|c|c|c|c|c|c|}
\hline
\multicolumn{2}{|c|}{\textbf{Geometry}} & \textbf{1}      & \textbf{2}     & \textbf{3}      & \textbf{4}\\ \hline
\multirow{7}{*}{\rotatebox[origin=c]{90}{\textbf{CV}}} & $N$                           & 52                  & 45                 & 35                  & 23 \\ \cline{2-6}
&$w$ (mm)                      & 0.475               & 0.438              & 0.369               & 0.263   \\ \cline{2-6}
&$B_{avg}$ (mT)                & 0.497               & 0.294             & 0.170               & 0.093    \\ \cline{2-6}
&$P_{\text{the}}$ (W)                              & 1.00                & 1.00               &1.00                 & 1.00		 \\ \cline{2-6}
&$P_{\text{sim}}$ (W)                              & 1.06                & 1.06               &1.05                 & 1.05	  \\ \cline{2-6}
&$m_{\text{the}}$ (Am$^2$)                       & 0.116                & 0.118               &0.112                 & 0.090	 \\ \cline{2-6}
&$m_{\text{sim}}$ (Am$^2$)                       & 0.122                & 0.124               &0.118                 & 0.094	  \\ \hline
\multirow{7}{*}{\rotatebox[origin=c]{90}{\textbf{CC}}} &$N$                           & 52                  & 45                 & 35                  & 23             \\ \cline{2-6}
&$w$ (mm)                              & 0.475               & 0.438              & 0.369              & 0.263          \\ \cline{2-6}
&$B_{avg}$ (mT)           	& 0.469               & 0.276             & 0.131               & 0.087       \\ \cline{2-6}
&$P_{\text{the}}$ (W)     		& 0.98                & 0.98               &0.98                 & 0.98          \\ \cline{2-6}
&$P_{\text{sim}}$ (W)  & 0.93                & 0.93               &0.93                 & 0.93\\ \cline{2-6}
&$m_{\text{the}}$ (Am$^2$)    			& 0.115                & 0.117               &0.111                 & 0.089          \\ \cline{2-6}
&$m_{\text{sim}}$ (Am$^2$)        			& 0.114                & 0.116               &0.111                 & 0.089          \\ \hline
\end{tabular}
\captionof{table}{Simulation geometry parameters and resultant averaged magnetic flux density as compared to their simulated values. Used in conjunction with those in Table \ref{ta:parameters}. }
\label{ta:results}
\end{table}
\end{multicols}

\begin{table}[H]
\begin{tabular}{|c|c|c|c|c|c|c|c|c|}
\hline
\textbf{Geometry}  & \textbf{$\bm{x}$ (m)} & \textbf{$\bm{y}$ (m)} & \textbf{$\bm{P}$ (W)} & \textbf{$\bm{V}$ (V)} & \textbf{$\bm{w}$ (mm)} & $\bm{N}$ & \textbf{$\bm{m}$ (Am$^2$)} & \textbf{Application}          \\ \hline
\multirow{4}{*}{1} & \multirow{4}{*}{0.1} & \multirow{4}{*}{0.1} & \multirow{2}{*}{1.0} & 3.3                  & 0.3465                & 61           & 0.0832                    & \multirow{4}{*}{1U, 2U, 3U, 6U} \\ \cline{5-8}
                   &                      &                      &                      & 5                    & -                     & -            & -                         &                                 \\ \cline{4-8}
                   &                      &                      & \multirow{2}{*}{2}   & 3.3                  & 0.5275                & 46           & 0.1272                    &                                 \\ \cline{5-8}
                   &                      &                      &                      & 5                    & 0.3200                & 66           & 0.1154                    &                                 \\ \hline
\multirow{6}{*}{2} & \multirow{6}{*}{0.2} & \multirow{6}{*}{0.1} & \multirow{2}{*}{0.5} & 3.3                  & 0.3349                & 68           & 0.1051                    & \multirow{6}{*}{2U, 3U, 6U}     \\ \cline{5-8}
                   &                      &                      &                      & 5                    & -                     & -            & -                         &                                 \\ \cline{4-8}
                   &                      &                      & \multirow{2}{*}{1.0} & 3.3                  & 0.5074                & 51           & 0.1612                    &                                 \\ \cline{5-8}
                   &                      &                      &                      & 5                    & 0.3071                & 72           & 0.1456                    &                                 \\ \cline{4-8}
                   &                      &                      & \multirow{2}{*}{2.0} & 3.3                  & 0.7582                & 38           & 0.2417                    &                                 \\ \cline{5-8}
                   &                      &                      &                      & 5                    & 0.4673                & 54           & 0.2246                    &                                 \\ \hline
\multirow{6}{*}{3} & \multirow{6}{*}{0.3} & \multirow{6}{*}{0.1} & \multirow{2}{*}{0.5} & 3.3                  & 0.4180                & 58           & 0.1454                    & \multirow{6}{*}{3U, 6U}         \\ \cline{5-8}
                   &                      &                      &                      & 5                    & 0.2552                & 83           & 0.1293                    &                                 \\ \cline{4-8}
                   &                      &                      & \multirow{2}{*}{1.0} & 3.3                  & 0.6239                & 43           & 0.2202                    &                                 \\ \cline{5-8}
                   &                      &                      &                      & 5                    & 0.3870                & 62           & 0.2022                    &                                 \\ \cline{4-8}
                   &                      &                      & \multirow{2}{*}{2.0} & 3.3                  & 0.9278                & 32           & 0.3272                    &                                 \\ \cline{5-8}
                   &                      &                      &                      & 5                    & 0.5791                & 46           & 0.3076                    &                                 \\ \hline
\multirow{8}{*}{4} & \multirow{8}{*}{0.3} & \multirow{8}{*}{2}   & \multirow{2}{*}{0.2} & 3.3                  & 0.3825                & 125          & 0.2188                    & \multirow{8}{*}{6U}             \\ \cline{5-8}
                   &                      &                      &                      & 5                    & -                     & -            & -                         &                                 \\ \cline{4-8}
                   &                      &                      & \multirow{2}{*}{0.5} & 3.3                  & 0.6566                & 85           & 0.3791                    &                                 \\ \cline{5-8}
                   &                      &                      &                      & 5                    & 0.4051                & 123          & 0.3489                    &                                 \\ \cline{4-8}
                   &                      &                      & \multirow{2}{*}{1.0} & 3.3                  & 0.9598                & 61           & 0.5621                    &                                 \\ \cline{5-8}
                   &                      &                      &                      & 5                    & 0.6053                & 90           & 0.5296                    &                                 \\ \cline{4-8}
                   &                      &                      & \multirow{2}{*}{2.0} & 3.3                  & 1.3954                & 44           & 0.8224                    &                                 \\ \cline{5-8}
                   &                      &                      &                      & 5                    & 0.8892                & 65           & 0.7881                    &                                 \\ \hline
\end{tabular}
\caption{Table of moments and optimal magnetorquer geometries given commonly used outer dimensions and a range of powers and voltages. Other parameters used are as follows: $s=0.254$ mm, $t=0.07$ mm, $\beta=0.9$, $\rho = 1.7\cdot 10^{-8}\SI{}{\Omega m}$. The CC model was not included in this table. Note that the CV magnetic moments are not scaled by the experimentally derived scaling of (1.05$\pm$0.02) (see Section \ref{Results-and-Discussion}).}
\label{ta:optimizations}
\end{table}

\end{document}